\documentclass[pra,preprint,superscriptaddress,amsmath,amssymb]{revtex4}
\usepackage{graphicx}% Include figure files
\usepackage{bm}
\usepackage{amsmath,amssymb}
\usepackage{color}
\include{Qcircuit}

\usepackage{theorem}

\theorembodyfont{\normalfont}
\newtheorem{theorem}{Theorem}
\newtheorem{definition}{Definition}

\begin{document}
YITP-19-19
\title{Impossibility of perfectly-secure one-round
delegated quantum computing for
classical client}
\author{Tomoyuki Morimae}
\email{tomoyuki.morimae@yukawa.kyoto-u.ac.jp}
\affiliation{Yukawa Institute for Theoretical Physics,
Kyoto University, Kitashirakawa Oiwakecho, Sakyo-ku, Kyoto 606-8502, Japan}
\affiliation{JST, PRESTO, 4-1-8 Honcho, Kawaguchi, Saitama 332-0012, Japan}
\affiliation{Department of Computer Science, 
Gunma University, 1-5-1 Tenjin-cho Kiryu-shi
Gunma-ken, 376-0052, Japan}
\author{Takeshi Koshiba}
\email{tkoshiba@waseda.jp}
\affiliation{Faculty of Education and Integrated Arts
and Sciences, Waseda University, Nishi-waseda 1-6-1, 
Shinjuku-ku, Tokyo 169-8050, Japan}

\begin{abstract}
Blind quantum computing protocols enable
a client, who can generate or measure
single-qubit states,
to delegate quantum computing to a remote quantum server 
protecting the client's privacy (i.e., input, output, and program).
With current technologies,
generations or measurements of single-qubit states
are not too much burden for
the client.
In other words, 
secure delegated quantum computing
is possible
for ``almost classical" clients.
However,
is it possible for a ``completely classical"
client?
Here we consider 
a one-round perfectly-secure delegated quantum computing,
and show that
the protocol cannot satisfy both
the correctness (i.e., the correct result is obtained
when the server is honest) and the perfect blindness (i.e.,
the client's privacy is completely protected)
simultaneously
unless BQP is in NP.
Since BQP is not believed to be in NP,
the result suggests the impossibility of 
the one-round perfectly-secure delegated quantum computing.
\end{abstract}

\date{\today}
\maketitle  

\section{Introduction}
Imagine that Alice who does not have any sophisticated quantum technology
wants to factorize a large integer.
She has a rich friend, Bob, who owns a full-fledged 
scalable quantum computer. 
Alice wants Bob to do the factoring for her.
However, the problem is that
Alice does not trust Bob, and therefore she does not want
to reveal her input, output, and the program (in this case
Shor's factoring algorithm) to Bob. 
Can she delegate her quantum computation to Bob while 
protecting her privacy?

\begin{figure}[htbp]
\begin{center}
\includegraphics[width=0.4\textwidth]{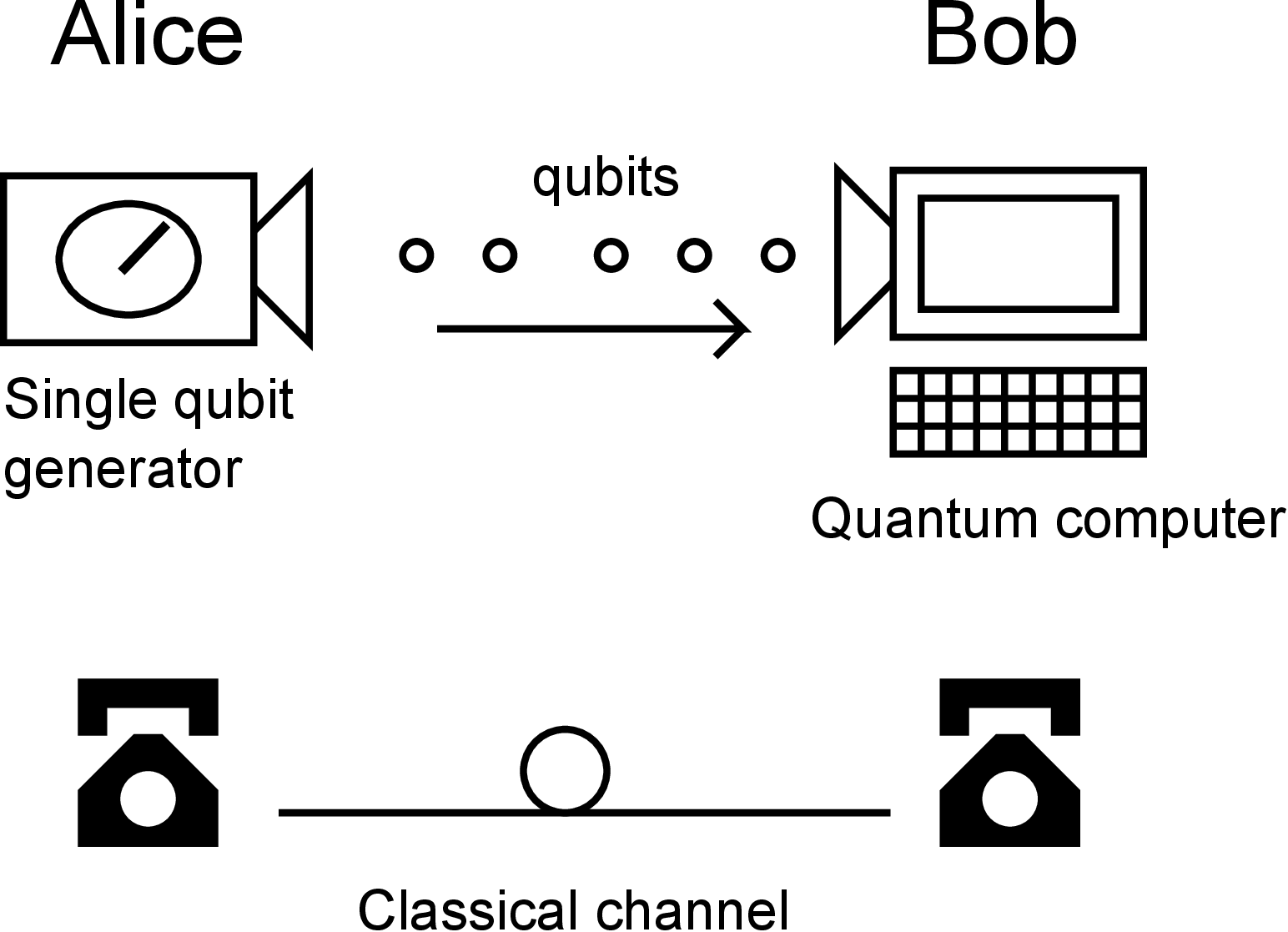}
\end{center}
\caption{
The blind quantum computing protocol proposed in Ref.~\cite{BFK}.
Alice possesses a device that emits randomly-rotated single-qubit states.
Bob has a universal quantum computer.
Alice and Bob share a two-way classical channel.
} 
\label{fig1}
\end{figure}

Broadbent, Fitzsimons, and Kashefi~\cite{BFK} theoretically showed
that such a secure delegated quantum
computing is indeed possible
if some minimum quantum technology is assumed for the client.
In the protocol of Ref.~\cite{BFK} (Fig.~\ref{fig1}),
Alice, a client, has  
a device that emits randomly rotated single qubit states. 
She sends these states to Bob, the server, who has the full quantum
technology.
Alice and Bob are also connected with a two-way classical channel.
Bob performs quantum computing by using qubits sent from Alice and 
classical messages exchanging with Alice via the classical channel.
After finishing his quantum computation, Bob sends the output of his computation,
which is a classical message, to Alice. 
This message encrypts the result of Alice's quantum computing, which is not
accessible to Bob.
Alice decrypts the message, and obtains the desired result of her quantum computing.
(Ref.~\cite{BFK} also proposed a quantum input and 
quantum output protocol.)
It was shown in Ref.~\cite{BFK} that whatever Bob does, he cannot learn
anything about the input, the program, and the output of Alice's 
computation
(except for some unavoidable leakage, such as upperbounds of
the sizes of the input, output, and program, etc.).
Proof-of-principle experiments were also done 
with photonic qubits~\cite{Barz,BarzNP,Chiara}.
The composable security of the protocol was also 
shown in Ref.~\cite{Vedrancomposability}.

In the protocol, the client has to possess a device that
generates single qubit states.
Generations of single qubit states are ubiquitous in today's laboratories,
and therefore not too much burden for the client.
In other words, ``almost classical" client can enjoy secure delegated quantum computing.

\begin{figure}[htbp]
\begin{center}
\includegraphics[width=0.4\textwidth]{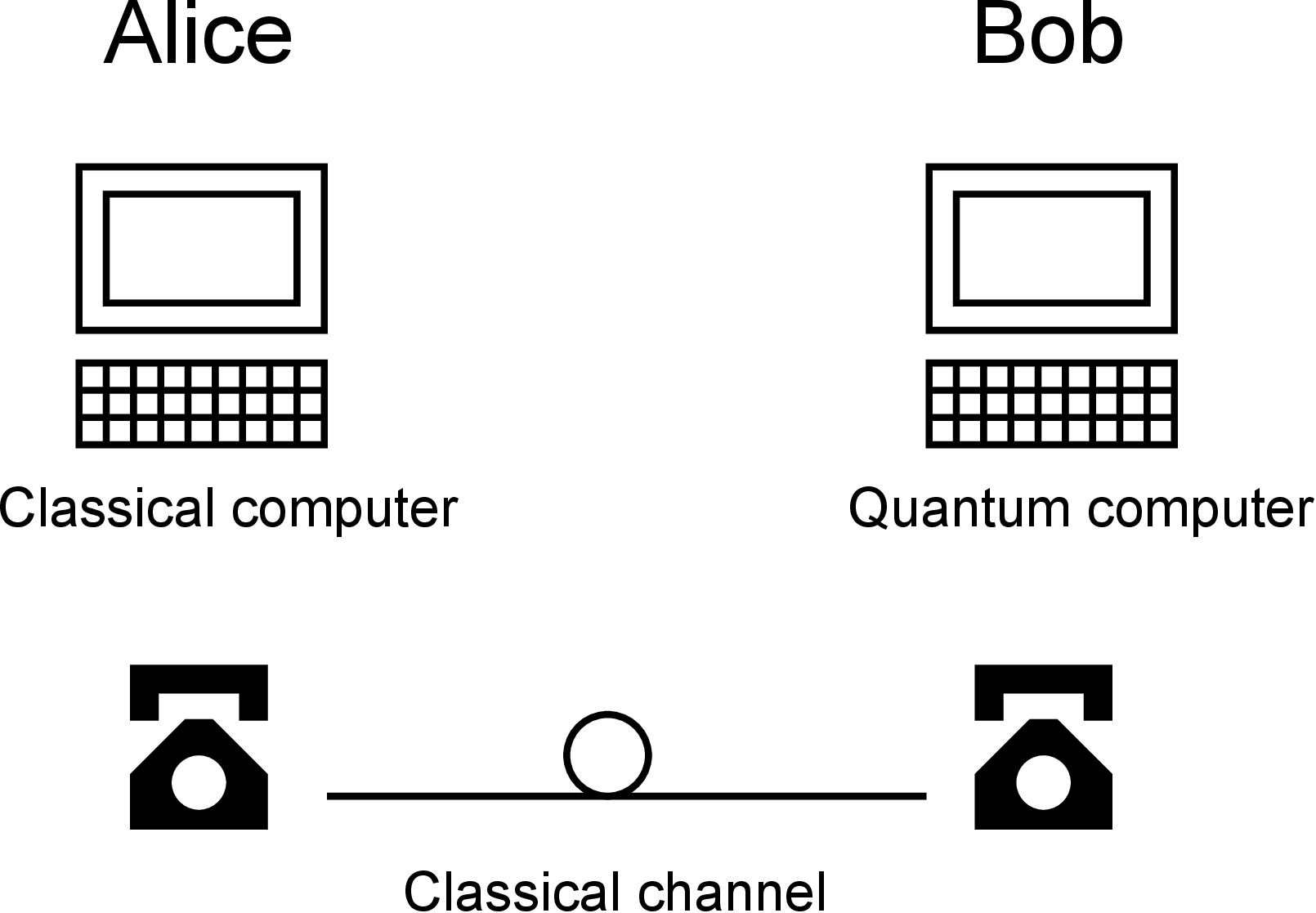}
\end{center}
\caption{
The secure delegated quantum computing for a classical client.
Alice has only a classical computer, whereas Bob has a universal quantum
computer.
Alice and Bob share a two-way classical channel.
} 
\label{classical}
\end{figure}

\begin{figure}[htbp]
\begin{center}
\includegraphics[width=0.4\textwidth]{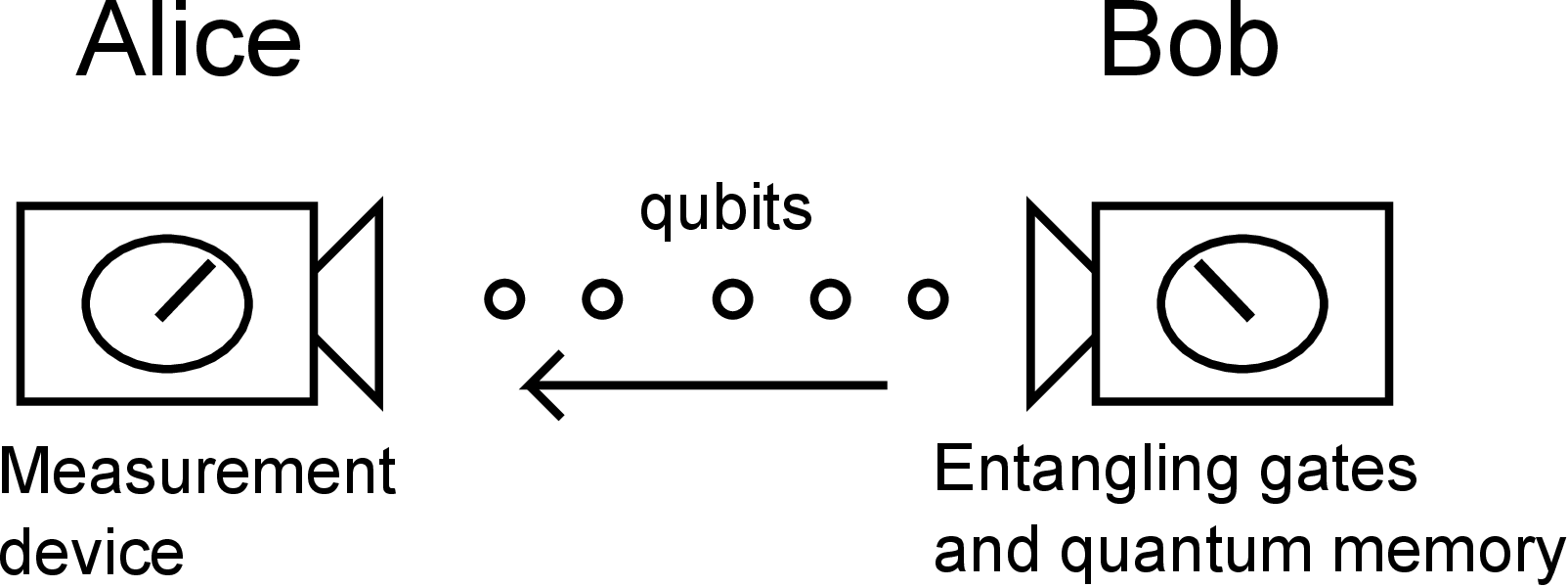}
\end{center}
\caption{
The blind quantum computing protocol proposed 
in Refs.~\cite{MABQC,HayashiMorimae}.
Alice possesses a device that measure qubits.
Bob has the ability of generating and storing
entangled many-qubit states.
} 
\label{measuringAlice}
\end{figure}

However, isn't it possible to realize secure delegated quantum computing
for a ``completely classical" client (Fig.~\ref{classical})?
Motivated by this question (and by other important questions such
as the verifiability~\cite{FK}),
many variant protocols for blind quantum computing have been 
proposed~\cite{
MABQC,FK,Vedran,AKLTblind,topoblind,CVblind,Lorenzo,
Joe_intern,Sueki,HayashiMorimae,Takeuchi,distillation,DunjkoKashefi}.
For example, it was shown that, instead of single-qubit states, 
the client has only to generate weak coherent pulse states
if we add more burden to the server~\cite{Vedran}.
Coherent states are considered as ``more classical" than single-photon states,
and therefore it enables secure delegated quantum computing for ``more classical"
client.
It was also shown that secure delegated quantum computing is possible
for a client who can only measure states~\cite{MABQC,HayashiMorimae} 
(Fig.~\ref{measuringAlice}). 
A measurement of a bulk state with a threshold detector is sometimes much easier than
the single-photon generation, and therefore the protocol also enables ``more classical"
client.
However, these protocols still require the client to have some minimum
quantum technologies,
namely the generation of weak coherent pulses or measurements of quantum states.
In fact, all protocols proposed so far require the client to have some 
minimum quantum abilities,
such as generations or measurements of quantum 
states.
(If we have two quantum servers,
a completely classical client can delegate quantum computing~\cite{BFK},
but in this case, we have to assume that two servers cannot communicate with each other.)

In short, the possibility of a 
perfectly-secure delegated quantum computing
for a completely-classical client is open.
(Note that the perfect security means that an encrypted text
gives no information about the plain text~\cite{nonlinearcrypto}.
It is a typical security notion 
in the information theoretical security.)

In this paper, we consider one-round perfectly-secure
delegated quantum computing for a completely-classical
client. We show that 
unless ${\rm BQP}\subseteq{\rm NP}$
it is impossible to
satisfy both the correctness 
and the blindness 
simultaneously (Theorem~\ref{theorem} below).
The definitions of the correctness and blindness are
given in Definition~\ref{def1} and Definition~\ref{def2} below,
respectively.
The containment of BQP in NP 
is not believed to happen~\cite{Watrous,BQPPH},
and therefore
the result suggests the impossibility of
one-round perfectly-secure delegated quantum computing
for a completely-classical client.

\section{Setup}
We first explain one-round perfectly-secure
delegated quantum computing for a completely-classical client.
Alice is completely classical, i.e., she
has only a probabilistic polynomial-time
Turing machine. 
Alice wants to solve a BQP problem.
In other words, she wants to decide whether $x\in L$ or $x\notin L$ for
an instance $x$ of a language $L$ in BQP.
However, Alice cannot do it by herself (unless 
${\rm BQP}={\rm BPP}$), and therefore
she delegates the computation to Bob as follows.
\begin{itemize}
\item[1.]
Alice generates a 
private key $k\in K$, where $K$
is the set of valid keys.
The key generation operation can be done in classical polynomial time.
We assume that checking the validness of a key 
can be done in classical polynomial time.
(Or, we assume that all bit strings are valid keys.)
She then encrypts $L$ and $x$ 
as $E_k(L,x)$, 
where $E$ is the encryption operation, which is deterministic
and in classical polynomial time.
She sends $E_k(L,x)$ to Bob.
%She also sends a polynomial $r\ge 2$ to Bob.
\item[2.]
Bob sends Alice 0 with probability $p_{Bob}(0|E_k(L,x))$
and 1 with probability 
$p_{Bob}(1|E_k(L,x))=1-p_{Bob}(0|E_k(L,x))$.
\item[3.]
Alice calculates 
the decrypting bit 
$d_k(L,x)\in\{0,1\}$, which can be calculated 
deterministically and in classical polynomial time. 
(It can be computed before she receives a bit from Bob.)
She accepts if and only if 
\begin{eqnarray*}
d_k(L,x)\oplus (\mbox{the bit sent from Bob})=1.
\end{eqnarray*}
\end{itemize}

When $d_k(L,x)=0$, Bob has to send 1 to make Alice accept.
On the other hand, if $d_k(L,x)=1$, Bob has to send 0 to make
Alice accept. In other words,
Bob's bit has to be equal to 
$d_k(L,x)\oplus 1$ to make Alice accept.
Therefore, for fixed $L$, $x$, and $k$, Alice's acceptance
probability $p_{Alice}(acc|L,x,k)$ is 
\begin{eqnarray*}
p_{Alice}(acc|L,x,k)=p_{Bob}(d_k(L,x)\oplus1|E_k(L,x)).
\end{eqnarray*}
We define the correctness and blindness as follows.
\begin{definition}
\label{def1}
[Correctness] 
We say that a protocol is correct if
the following is satisfied.
For any language $L\in{\rm BQP}$, instance $x$,
and private key $k\in K$, 
%and polynomial $r\ge2$, 
if $x\in L$ then 
\begin{eqnarray*}
p_{Alice}(acc|L,x,k)\ge c, 
\end{eqnarray*}
while if
$x\notin L$ then 
\begin{eqnarray*}
p_{Alice}(acc|L,x,k)\le s,
\end{eqnarray*}
where $c>\frac{1}{2}$,
$0\le s<c\le1$, and $c-s\ge1/poly(|x|)$. 
\end{definition}
\begin{definition}
\label{def2}
[Blindness] Informally, blindness means that
Bob cannot learn anything about Alice's $(L,x)$
from $E_k(L,x)$.
More formaly, we say that an encryption is blind if the
following is satisfied. For any $L_1,L_2\in {\rm BQP}$,
valid key $k_1$,
$x_1\in L_1$, and $x_2\in L_2$, 
there exists a valid key $k_2$ such that
\begin{eqnarray*}
E_{k_1}(L_1,x_1)=E_{k_2}(L_2,x_2).
\end{eqnarray*}
\end{definition}

Note that the above delegation protocol is not the most general one.
First, the encryption operation by Alice is deterministic and symmetric.
It is open whether we can consider more generalized encryptions.
Second, Bob sends only a single bit of message to Alice.
(Regarding this point, see the Discussion section.)
Finally, Alice's decryption operation is not the most general one.

\section{Result}
Now we show our main result:

\begin{theorem}
\label{theorem}
If the above protocol
satisfies both the correctness and blindness 
simultaneously,
then ${\rm BQP}\subseteq{\rm NP}$.
\end{theorem}

{\it Proof}.---
Let $L$ be a language in BQP.
We show that the following NP protocol can verify $L$.
\begin{itemize}
\item[1.]
Merlin sends polynomial-length classical bit strings
$w$ and $w_0$ to Arthur.
If Merlin is honest, $w_0$ is any private key from $K$,
and $w$ is a key from $K$ that satisfies
\begin{eqnarray}
E_{w_0}(L_0,0)=E_w(L,x),
\label{equal}
\end{eqnarray}
where 
\begin{eqnarray*}
L_0\equiv\{x\in\{0,1\}^*~|~\mbox{the first bit of $x$ is 0}\}.
\end{eqnarray*}
Obviously, $0\in L_0$ and $L_0\in {\rm BQP}$.
Note that such $w$ always exists for any $w_0$, since otherwise
Bob can learn that Alice's computation is not $(L,x)$ when he receives
$E_{w_0}(L_0,0)$, which contradicts the blindness.
\item[2.]
Arthur checks whether $w$ and $w_0$ are valid keys.
(We have assumed that the check can be done in classical polynomial time,
or all bit strings are valid keys.)
If at least one of them is invalid, Arthur rejects and
aborts.
\item[3.]
Arthur calculates $E_w(L,x)$ and $E_{w_0}(L_0,0)$,
which can be done deterministically and in classical polynomial time.
Arthur rejects and aborts if 
\begin{eqnarray*}
E_w(L,x)\neq
E_{w_0}(L_0,0).
\end{eqnarray*}
%\item[3.]
%Arthur rejects and aborts if
%\begin{eqnarray*}
%w_3\oplus d_{w_1}=1.
%\end{eqnarray*}
\item[4.]
Arthur calculates $d_w(L,x)$ and $d_{w_0}(L_0,0)$, which can be done
deterministically and in classical polynomial time. Arthur accepts if and only if
\begin{eqnarray*}
d_w(L,x)=d_{w_0}(L_0,0).
\end{eqnarray*}
\end{itemize}

We show that this NP protocol 
can verify $L$.
Note that due to the correctness,
\begin{eqnarray}
p_{Bob}(d_k(L_0,0)\oplus 1|E_k(L_0,0))\ge c
\label{co1}
\end{eqnarray}
for any key $k\in K$.
%and any polynomial $r\ge2$.

First let us consider the case of
$x\in L$. 
In this case, due to the correctness,
\begin{eqnarray}
p_{Bob}(d_k(L,x)\oplus 1|E_k(L,x))\ge c
\label{co2}
\end{eqnarray}
for any key $k\in K$.
%and any polynomial $r$.
Furthermore, Arthur never rejects at steps 2 and 3.
%Furthermore, Arthur never rejects at step 3, since
%\begin{eqnarray*}
%Pr[w_3\oplus d_{w_1}=0]
%&=&
%Pr[Q(E(I^{\otimes n},k_0))\oplus d_{k_0}=0]\\
%&=&1.
%\end{eqnarray*}
%Finally, 
Finally, we can show $d_w(L,x)=d_{w_0}(L_0,0)$ and therefore
Arthur accepts.
In fact,
if $d_w(L,x)\neq d_{w_0}(L_0,0)$,
which means 
\begin{eqnarray}
d_{w_0}(L_0,0)=d_w(L,x)\oplus1,
\label{d}
\end{eqnarray}
then
\begin{eqnarray*}
c&\le&
p_{Bob}(d_{w_0}(L_0,0)\oplus1|E_{w_0}(L_0,0))
~~~(\mbox{from Eq.~(\ref{co1})})\\
&=&p_{Bob}(d_{w_0}(L_0,0)\oplus1|E_w(L,x))
~~~(\mbox{from Eq.~(\ref{equal})})\\
&=&p_{Bob}(d_w(L,x)|E_w(L,x))~~~(\mbox{from Eq.~(\ref{d})})\\
&=&1-p_{Bob}(d_w(L,x)\oplus1|E_w(L,x))\\
&\le&1-c~~~(\mbox{from Eq.~(\ref{co2})}),
\end{eqnarray*}
which contradicts to $c>\frac{1}{2}$. Therefore, 
Arthur accepts when $x\in L$.

Next let us consider the case of $x\notin L$.
In this case, due to the correctness,
\begin{eqnarray}
p_{Bob}(d_k(L,x)\oplus 1|E_k(L,x))\le s
\label{co3}
\end{eqnarray}
for any key $k\in K$.
%and any polynomial $r$.
If Arthur arrives at step 4,
$w$ and $w_0$ are valid keys, and
\begin{eqnarray}
E_{w}(L,x)=E_{w_0}(L_0,0)
\label{equal2}
\end{eqnarray}
is satisfied.
Let us assume that
\begin{eqnarray}
d_w(L,x)=d_{w_0}(L_0,0).
\label{d2}
\end{eqnarray}
Then,
\begin{eqnarray*}
c&\le&
p_{Bob}(d_{w_0}(L_0,0)\oplus1|E_{w_0}(L_0,0))
~~~(\mbox{from Eq.~(\ref{co1})})\\
&=&p_{Bob}(d_{w_0}(L_0,0)\oplus1|E_w(L,x))
~~~(\mbox{from Eq.~(\ref{equal2})})\\
&=&p_{Bob}(d_w(L,x)\oplus1|E_w(L,x))~~~(\mbox{from Eq.~(\ref{d2})})\\
&\le&s~~~(\mbox{from Eq.~(\ref{co3})}),
\end{eqnarray*}
which contradicts to $s<c$. Therefore, 
$d_w(L,x)\neq d_{w_0}(L_0,0)$,
which means that Arthur rejects.
In summary, we have shown that $L$ is in NP.

\section{Discussion}
In this paper, we have shown that unless ${\rm BQP}\subseteq{\rm NP}$
one-round perfectly-secure delegated quantum computing
cannot satisfy both
the correctness and the perfect blindness 
simultaneously.

If we relax the requirement of the perfect security
to a computational one, for example,
there would be several ways of secure delegated quantum computing
for a classical client~\cite{Gentry,Mahadev,Cojocaru,Brakerski}.
For example, 
the fully-homomorphic encryption scheme~\cite{Gentry}
might be able to achieve secure delegated quantum computing
for a classical client.
Recently, a secure delegated quantum computing protocol
for a completely classical client has been proposed
by using the Learning With Errors problem~\cite{Mahadev}.

In our proof, we do not assume $c-s\ge 1/poly$.
Therefore, a similar proof shows that
if PP can be solved in the protocol,
then the polynomial hierarchy collapses.

Finally, we point out that a related result was obtained in Ref.~\cite{Yu},
where an impossibility result of an information-theoretically-secure
quantum homomorphic encryption
was derived by showing that the size of the sending message
from Alice to Bob must be exponentially large to hide polynomial-size
quantum circuits.
We also mention that after uploading the first version of this
paper on arXiv,
more general results on the impossibilities of
secure delegated quantum computing with a completely classical client
have been obtained~\cite{Aaronson}.
In particular, Ref.~\cite{Aaronson} considers more general
case where polynomial-length messages are exchanged in polynomial-round
between the server and the client, while the present paper
considers the limited case where only a single bit is sent from
the server to the client.
On the other hand, the complexity conjecture,
${\rm BQP}\not\subseteq{\rm NP}$,
that our result is based on
has an oracle separation~\cite{Watrous,BQPPH},
while 
that of Ref.~\cite{Aaronson},
${\rm BQP}\not\subseteq{\rm NP}/{\rm poly}\cap{\rm coNP}/{\rm poly}$,
does not~\cite{Aaronson}.
It is not clear how to generalize our result to the more general
case where the server sends a polynomial-length bit string
without introducing advice.

We also mention that Refs.~\cite{Aaronson,Msampling} consider delegations
of sampling of output probability distributions of sub-universal
quantum computing models, while here we consider delegations of
decision problems in BQP, which does not seem to be directly applied
to the sampling.

\acknowledgements
TM is supported by MEXT Q-LEAP, JST PRESTO No.JPMJPR176A and 
JSPS Grant-in-Aid for Young Scientists (B)
No.JP17K12637.
TK is supported by JSPS Grant-in-Aid for Scientific Research (A)
JP16H01705 and for Scientific Research (B) JP17H01695.

\end{document}